# Multiparametric qualimetric microsurgical scanning chip-lancet model: theoretical metrological and biomedical considerations

A. Jablokov[1], Oleg Gradov[2,3]*

[1]Russian National Research Medical University, Moscow Faculty, Moscow, Russia; [2]Institute of Biology and Chemistry (MPSU), Department of Anatomy and Physiology of Humans and Animals, Moscow, Russia; [3]Institute of Energy Problems of Chemical Physics, Russian Academy of Sciences, Moscow, Russia.
* Corresponding author: Oleg Gradov, Lelninsky pr., 38, bld. 2, Cab. 18, Moscow, 119334, Russia;
e-mail: o.v.gradov@gmail.com; neurobiophys@gmail.com

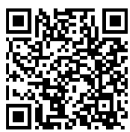

**ABSTRACT**

The construction of a novel surgical instrument is considered, which is also a probing device providing a signal to the measuring equipment, which after its interpretation allows to obtain useful information about the section quality and the biomaterial properties. We propose here some formalized considerations on the possibility of its implementation for different variables registration. The idea is also extrapolated into the field of micrurgy which refers to the microelectrode techniques and the local potential registration in situ.

## 1. INTRODUCTION

Let us assume that there is a scalpel/lancet combined with a detecting head (see Fig. 1) with the known transfer function in the diagnostic spectral range. Spectral distribution pattern, that is a position sensitive spectral data frame detected at the output of the tissue, i.e. at the detecting device input, depending on the changeable chemism of the tissue (flap) points, has the form $\Theta_{in}(x, y, t, \lambda)$, where $(x, y)$ are the plane coordinates, $t$ - time and $\lambda$ - wavelength. This is true for both optical and mass spectra, but in the latter case mass to charge ratio or the ion current intensity dependence on the mass to charge ration $M$ instead of $\lambda$ is detected. The presence of the time variable $t$ allows the flap or a tissue fragment diagnostics within dynamic chemistry, since the function dimension for a statistic pattern is reduced by one to the function of three variables $\Theta_{in}(x, y, \lambda)$ or $\Theta_{in}(x, y, M)$.

The sample scanning obviously includes a coordinate transformation at a scale equal to the moving / scanning rate of a spectrometric or a mass spectrometric head relative to the movement of the non-fixed tissue flap in the course of the operation - the analytical signal deconvolution. Similar to the discretization with respect to the argument $\lambda$ by considering not the entire set of the $\Psi(\lambda)$ function values, describing the optical emission spectrum of the sample area, but only its additive RGB ranges $\Psi_R(x,y,t)$, $\Psi_G(x,y,t)$, $\Psi_B(x,y,t)$, used in multispectral digital photography, in the case of a mass-spectrometric control it is expedient to analyze only the most interesting from the diagnostic point of view mass-spectral distribution areas with different masses. Suppose that we know that a diagnostic agent for certain a disease *N* belongs to the mass range *N*, while the diagnostic agent for another disease *M* lies in the *M* mass range. Than it is possible to divide the **M** area into two regions $\Theta_N(x,y,t,\mathbf{M})$ and $\Theta_M(x,y,t,\mathbf{M})$ for determination of the particular diagnostic agent of interest (see Fig. 2 and 3). This reduces the data aquisition representativeness but significantly simplifies the procedure and increases the rate of the intraoperative agent distinction.

scanning mechanism. In the case of the analysis at the exactly known wavelengths or molecular fragment mass ranges we can write

$$\Theta_{N_1}(x,y), \Theta_{N_2}(x,y), \ldots, \Theta_{N_Z}(x,y),$$
$$\Theta_{M_1}(x,y), \Theta_{M_2}(x,y), \ldots, \Theta_{M_Z}(x,y)$$

where **Z** is a set of integers (full sampling of the spectral frame, which can be rewind either forward or backward for the corresponding **Z** which is in principle unlimited). In a general case for any arbitrary wavelength $\lambda$ or any mass **M**

$$\Theta_{\square\lambda_1}(x,y), \Theta_{\square\lambda_2}(x,y), \ldots, \Theta_{\square\lambda_Z}(x,y)$$
$$\Theta_{\square M_1}(x,y), \Theta_{\square M_2}(x,y), \ldots, \Theta_{\square M_Z}(x,y)$$

where $\forall \lambda \in \square$ and $\forall \mathbf{M} \in \square$. In other words, the data frame sequence for spectroscopic or mass-spectrometric image-guided surgery can be obtained either from single masses or diagnostic wavelengths, or from the wide range spectral distributions.

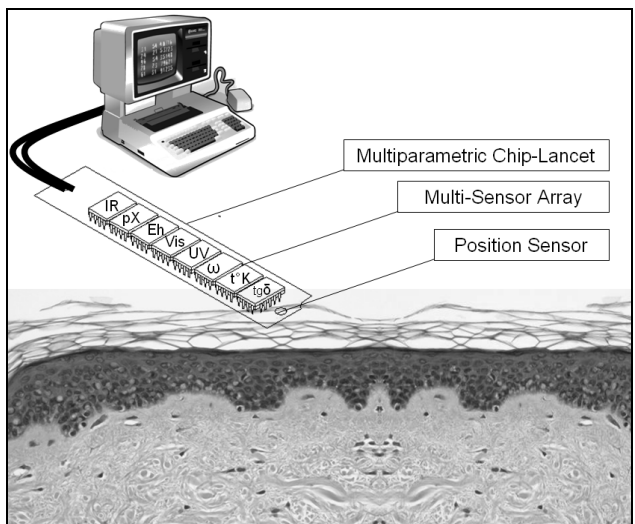

**Fig. 1.** The component scheme of the Multiparametric Chip-Lancet for epitelial surgery.

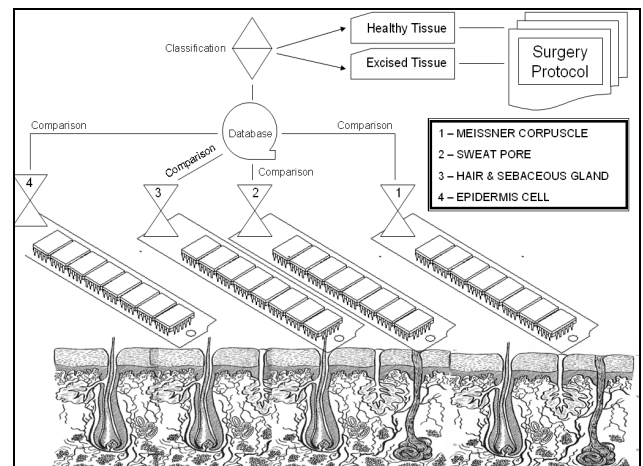

**Fig. 2.** Example of the intraoperative multiparametric analytical signal classification from derma.

## 2. PHENOMENOLOGICAL MODEL

For the dynamic control it is necessary to consider discretization with respect to the argument $t$, performed by the signal recording at the intervals corresponding to the frame changing periods. Herewith the four-dimensional function describing the signal transforms (is converted) into a sequence of two-dimensional functions due to the spectral

Let us now introduce an operator $\varphi[\ldots]$ that shows the way we should impact the function determining the input signal in order to obtain the output signal function. When the detection is performed along the scalpel movement line the signal which forms the data frame is one-dimensional and can be expressed by the function $\Theta(\vartheta)$, where $\vartheta$ stands for the spatial coordinates, time and wavelengths (or masses detected), besides the input and output functions are related as $\Theta_{out}(\vartheta) = \varphi[\Theta_{in}(\vartheta)]$. Then it is obvious that the input signal of the registration device during the analyte sampling in a mass-spectrometric scalpel or during the optical gap opening in the optical-spectroscopic control can be expressed by the delta-function $\Theta_{in}(\vartheta) = \delta[\vartheta - \vartheta_1]$, where $\vartheta_1$ stands for the pulse coordinate at the input. It follows that the impulse response function at the output is $\Theta_{out}(\vartheta, \vartheta_1) = \varphi[\delta(\vartheta - \vartheta_1)]$, resulting in a limited temporal resolution which correlates with the limited spatial intervals at the line of resection and registration when they are synchronized. If the system is linear, it satisfies the superpositoin principle. Thus, if we connect the 2D-input and output signals by the ratio $\Upsilon_{out}(x,y) = \varphi[\Upsilon_{in}(x,y)]$, then

$$\varphi\left[\sum_m a_m \Upsilon_{in_m}(x,y)\right] = \sum_m a_m \varphi\left[\Upsilon_{in_m}(x,y)\right].$$

Suppose further that during the analyte sampling we fed an infinitesimal spatial impulse, characterized by the delta-function $\Upsilon_{in}(x,y) = \delta(x - x_1, y - y_1)$, where $x_1, y_1$ stand for the spatial impulse coordinates at the input of the detection system, to the input of the spectral imaging registration device. The impulse dissipation and scattering occurs in the course of the detection process, therefore it is better to consider the point spread function rather then the point of analysis, which can be denoted $\Upsilon_\tau$

$$\Upsilon_{out}(x,y; x_1,y_1) = \Upsilon_\tau(x,y; x_1,y_1) = \varphi[\delta(x-x_1, y-y_1)]$$

As a result of the input signal decomposition into several points located at short distances at the scalpel trajectory $X_1$ along the $0x$ axis and $Y_1$ along $0y$ axis we can obtain a unified expression for the input signal:

$$\Upsilon_{in}(x,y) \cong \sum_n \sum_k \Upsilon_{in}(nX_1, kY_1) \delta(x-nX_1, y-kY_1) X_1 Y_1$$

and, therefore,

$$\Upsilon_{out}(x,y) \cong \varphi\left[\sum_n \sum_k \Upsilon_{in}(nX_1, kY_1) \delta(x-nX_1, y-kY_1) X_1 Y_1\right]$$

and

$$\Upsilon_{out}(x,y) \cong \sum_n \sum_k \Upsilon_{in}(nX_1, kY_1) \Upsilon_\tau(x,y; nX_1, kY_1) X_1 Y_1$$

It is also advisable to replace the sums in the above expression with the double integral:

$$\Upsilon_{out}(x,y) = \iint_{-\infty}^{\infty} \Upsilon_{in}(x_1, y_1) \Upsilon_\tau(x,y; x_1, y_1) dx_1 dy_1$$

Assuming that the scalpel and the sampler are always located at the same angle to the fixed operated flap, i.e. the sampler or the optical gap projections remain isoplanar and do not change their shape during the scanning procedure, one can obtain a two-dimensional convolution of the input function with the point spread function:

$$\Upsilon_{out}(x,y) = \iint_{-\infty}^{\infty} \Upsilon_{in}(x_1, y_1) \Upsilon_\tau(x-x_1, y-y_1) dx_1 dy_1.$$

For the linear and isoplanar (i.e. spatially invariant) registration system at the qualimetic scalpel its properties can be completely determined from the point spread function.

## 3. IF THE RESECTION IS PERFORMED ALONG A STRAIGHT LINE…

If the resection is performed along a straight line, parallelizable with the positional alignment axis of the surgical instrument (e.g. along the $0y$ axis), leading to the simplification of the expression $\Upsilon_{in}(x,y) = \delta(x)$, then in case of the isotropic circular symmetry of the point spread function we can determine the analyte distribution in the output signal along the resection axis. For the specified function with the circular symmetry:

$$\Upsilon_{out}(x,0) = \iint_{-\infty}^{\infty} \Upsilon_\tau(x_1, y_1) \delta(x - x_1) dx_1 dy_1$$

or, more briefly:

$$\Upsilon_{out}(x,0) = \int_{-\infty}^{\infty} \Upsilon_\tau(x,y) dy_1 \ .$$

If we replace $y_1 :\Leftrightarrow y$ that is rather possible in general case, we introduce a line spread function:

$$\Upsilon_{line}(x) = \Upsilon_{out}(x,0) = \int_{-\infty}^{\infty} \Upsilon_\tau(x,y) dy$$

If the system possesses a point spread function with separable variables, i.e.

$$\Upsilon_\tau(x,y) = \Upsilon_{\tau x}(x) \Upsilon_{\tau y}(y),$$

then:

$$\Upsilon_{line}(x) = \int_{-\infty}^{\infty} \Upsilon_{\tau x}(x) \Upsilon_{\tau y}(y) dy = \Upsilon_{\tau x}(x) \int_{-\infty}^{\infty} \Upsilon_{\tau y}(y) dy$$

while another is obvious for isotropic systems:

$$\Upsilon_{line}(x) = \int_{-\infty}^{\infty} \Upsilon_\tau(r) dy = \int_{-\infty}^{\infty} \Upsilon_\tau(\sqrt{x^2 + y^2}) dy$$

where $r$ is the focusing radii of the analyzed spot.

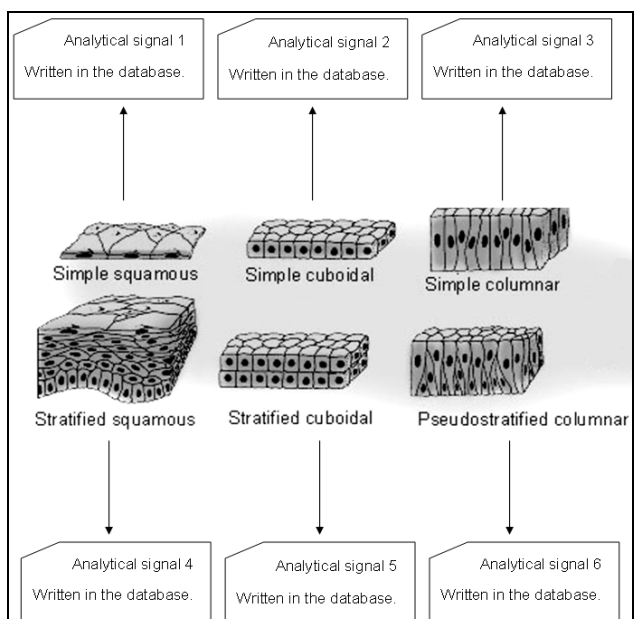

**Fig. 3.** Another example of the tissue signal classification for comparative intraoperation analysis followed by the control of the surgical instrument positioning accuracy.

## 4. CONCLUSION

If we consider each discrete cell (point) of such a measurement as a part of the detection matrix of a particular descriptor or correlation predictor, then the positioning accuracy and the correspondence between the resection discretization steps and the registration steps can be regarded as a correctness criterion of the position-sensitive analyte detection. In this connection the cells must not overlap and the analytical signals from the tissue must not interfere. However, the point spread functions extending beyond the boundaries of a single cell and leading to the correlation between the adjacent measuring positions or the memory cells/units, reduce the information capacity of the whole analytical system and the determination accuracy of the significant biochemical components, that must be taken into account during surgical qualimetry. The model has been detailed in a recent Russian publication [1] and short English conference version [2].

A comprehensive review of microsurgical scanning qualimetry is presented by Smith et al. [3, 4]. The justification for this solution scheme is the method for hybridization and complexation of different descriptors [5, 6] on a scanning line dataflow. This approach is best illustrated by an example of the *in situ* microbeam-assisted positional-sensitive measurements of the tissue parameters during the microsurgery (micrurgy) manipulations on the living brain slices [7].

## AUTHORS' CONTRIBUTION

OG: Mathematical conception, principal construction of the chip-lancet/chip-scalpel and writing of the manuscript; AJ: Micro-medical applicability realization and technical approbation of the concept. Both authors read and approved the final manuscript.

## TRANSPARENCY DECLARATION

The authors declare no conflicts of interest.